\documentclass[review]{elsarticle}
\graphicspath{ {./figures/} }
\usepackage{hyperref}
\usepackage{float}
\usepackage{verbatim} %comments
\usepackage{apalike}
\restylefloat{figure}
\restylefloat{table}
\usepackage{amsmath,amssymb,amsfonts}
\usepackage{graphicx}
\usepackage{xcolor}
\def\BibTeX{{\rm B\kern-.05em{\sc i\kern-.025em b}\kern-.08em
    T\kern-.1667em\lower.7ex\hbox{E}\kern-.125emX}}
\usepackage{lineno}
\usepackage{multirow}
\usepackage{caption}
\usepackage{subcaption}
\usepackage{mathdots}
\usepackage[classicReIm]{kpfonts}
\usepackage[english]{babel}
\usepackage{enumitem}
\usepackage{algorithm}
\usepackage{algorithmic}
\setlist{nolistsep}

\journal{Expert Systems with Applications}

\biboptions{authoryear}

\begin{document}
\begin{frontmatter}

\title{Medical Information Retrieval and Interpretation: A Question-Answer based Interaction Model}

\author[label1]{Nilanjan Sinhababu}
\ead{nilanjansb95@gmail.com}

\author[label2]{Rahul Saxena}
\ead{rahulsaxena31iitkgp@gmail.com}

\author[label1]{Monalisa Sarma}
\ead{monalisa@iitkgp.ac.in}

\author[label3]{Debasis Samanta \corref{cor1}}
\ead{debasis.samanta.iitkgp@gmail.com}

\cortext[cor1]{Corresponding author.}
\address[label1]{Subir Chowdhury School of Quality and Reliability, IIT Kharagpur, India}
\address[label2]{Department of Chemical Engineering, IIT Kharagpur, India}
\address[label3]{Department of Computer Science and Engineering, IIT Kharagpur, India}

\begin{abstract}
The Internet has become a very powerful platform where diverse medical information are expressed daily. Recently, a huge growth is seen in searches like symptoms, diseases, medicines and many other health related queries around the globe. The search engines typically populates the result by using the single query provided by the user and hence reaching to the final result may require a lot of manual filtering from the user's end. Current search engines and recommendation systems still lacks real time interactions that may provide more precise result generation. This paper proposes an intelligent and interactive system tied up with the vast medical big data repository in the web and illustrates its potential in finding medical information.
\vspace{1cm}
\end{abstract}


\end{frontmatter}

%% main text
\section{Introduction}
\label{S:1}

\subsection{Context}
\label{S:1.1}
Numerous health related information are streamlined by the Internet. The Internet empowers users to pick up fast admittance to data that can help in the analysis of medical issues or the advancement of appropriate treatments. It enable consumers to gather health-related information themselves and from the comfort of their home. The Internet can likewise provide various health related information beyond the immediate arrangement of care. It requires no administrative or financial overheads. Provided these benefits, many consumers tend to use internet as a source of their medical information. But at the end of the day, searching through this huge amount of data and collecting health related information is quite difficult of not impossible task. Also, there is always a chance of mistake in manual checking of the results provided by the search engines. Although with advancements in search engine technologies the searching is far superior than before, but they are built to provide immediate results based on a single query. Hence, these kind of systems are only suitable for advanced users at least for medical domains. The process of finding a information in internet can be divided into two major steps. Firstly, the extraction of data present in the web regarding a particular query provided by the user. This is the step that the search engines are well optimised to do. Secondly, narrowing down the huge data collected in the extraction phase by using some interaction mechanism. And finally, providing the user with the well filtered data that does not require manual filtering from the user's end. To perform such kind of intelligent searching, the two most important techniques required are information retrieval and natural language interactive question generation. Information discovery and interaction systems have been closely connected for many years and are the two aspects of computation that has proven to be superior in various domains. Recent advances in these techniques have provided a way of enabling truly “intelligent” medical information retrieval system. \par

Presence of huge volume of data, specifically unstructured data in the internet has made it a quite difficult task to discover and find crucial information. Information retrieval is a key in many fields with an intent to collect information or develop knowledge databases. Earlier, these unstructured information required manual intervention for any kind of knowledge discovery. But with advancement of machine learning techniques the process of extracting information has become automatic and hence gained a lot of importance currently.  With a lot of research in the machine learning for text data, various methods for information discovery has been proposed. Techniques such as embedding and clustering has provided the community with ways of dealing with unlabeled text data.\par

On the other hand, question answer modelling systems are becoming one of the most important and emerging areas of interest in the research community. Earlier interaction with a computer were limited to rule based methods only. But with improvements in deep learning models, computer interactions are becoming more intelligent and accurate. These systems are able to help teachers and students by generating intelligent questions like fill in the blanks, MCQs and subjective questions from paragraphs, sentences and words. Furthermore, these systems are already proved to be beneficial to the artificial intelligence assistants and robots where interaction is required from both parties \cite{HCI-005}. Question answering system can be beneficial for some casual users who may ask simple factual questions\cite{heilman2011automatic}, Doctors who seek quick answers on medicine or medical equipment; Patients who seek medical treatment information online\cite{saito1988medical}; This paper presents a novel way of extracting unstructured information from the internet and providing users with medical queries the most useful results using interaction. 

\subsection{State of the Art}
\label{S:1.2}
There are various research works that focus on converting unstructured data into some structured form for information retrieval purpose \cite{mccallum2005information, augenstein2012lodifier, zainol2018visualurtext}. They generally use rule based methods for known data and for unknown data general machine learning and statistical methods are applied to get certain information\cite{mccallum2005information, ahern2007world, nesi2014ge}. Regarding the question generation models, the existing techniques greatly depend on the context of the input text documents and the required outputs. General RNN models does not capture enough context information to generate grammatically and contextually correct questions \cite{du2017learning}. To overcome this issue, LSTM models are proposed and bi-LSTM in general performs the best considering the problem of question generation. To further improve the question generation models, attention and coverage mechanisms are utilized, outperforming other models in this domain \cite{du2017learning, chali-baghaee-2018-automatic}.

\subsection{Motivation and Scope}
\label{S:1.3}
Existing information retrieval methodologies are prone to unreliability due increasing variety of enormous unstructured data\cite{INR-012}. These models are generally static and serves only for a specific type of data \cite{mccallum2005information, nesi2014ge}. With our literature survey, we were unable to find a research work that focuses on dynamic information retrieval. Further, information retrieval by reducing the unstructured web data corpus based on a interaction mechanism is totally missing in the literature. Our objective is to dynamically filter data to provide users with the minimal focused data using some interaction mechanisms. To do this work, the existing works for information retrieval are not sufficient. Further, bi-LSTM with attention and coverage suffers the most when the input sequence is too small \cite{du2017learning}. That means, the model performance is greatly dependent on the length of the input sequences provided. This particular work requires the question generation model to be able to generate a valid context based question even with a single word, but regarding the current state of the art techniques, they are prone error for smaller sequences of input. A model that can generate proper context based questions even with small (single word) of the input sequences is missing and needs to be investigated.

\subsection{Objectives}
\label{S:1.4}
The importance of question generation models are taken into consideration to address the man-machine interaction problem and how some intelligent question(s) can be formulated in order to retrieve some relevant context information. Further, these context information can be interpret using the answer(s) provided by user to derive some decision. To overcome limitation of the existing models, this research paper has three main objective as pointed below:
\begin{enumerate}
  \item Identifying a questionable sentence, which have enough context information for generating some decision.
  \item Generating contextually correct and medical related natural language questions from the questionable entity even with smaller sequences of inputs.
  \item  Analysing the answers provided by the user to reduce the corpus volume.
\end{enumerate}
\vspace{0.5cm}
\par The later section of this paper is arranged in the following sequence. First, the related work and state of the art is identified. Then the task deﬁnition is outlined, followed by the demonstration of the model structure. Then the experimental parameters are discussed for the experiment and finally, results, discussion and validity of the model is presented.

\section{Related Work}
\label{S:2}
In this section, the related work is divided into different sections using the different categories that the documents contain. This can be helpful in categorization of the content of the research papers more easily and hence the readability is highly increased. The related work is divided for information retrieval in unstructured data and question answer generation techniques.

\subsection{Information retrieval in unstructured data}
\label{S:2.1}

Andrew McCallum in 2005 \cite{mccallum2005information} showed different ways to convert raw unstructured data from web search to structured data were discussed like rule based model, hidden Markov model,  conditional probability models, text classification etc. 

Johanna Fulda et al. in 2015 \cite{fulda2015timelinecurator} developed a browser-based authoring tool that automatically extracts event data from temporal references in unstructured text documents using natural language processing and encodes them along a visual timeline. It uses context-dependent semantic parsing for entity extraction.

Isabelle Augenstein et al. in 2012 \cite{augenstein2012lodifier} modelled an approach (LODifier) that combines deep semantic analysis with named entity recognition, word sense disambiguation and controlled Semantic Web vocabularies in order to extract named entities and relations between them from text and to convert them into an RDF representation which is linked to DBpedia and WordNet.

Rabeah Al-Zaidy et al. in 2012 \cite{al2012mining} provide an end-to-end solution to automatically discover, analyze, and visualize criminal communities from unstructured textual data. Discover all prominent communities and measure the closeness among the members. generate all indirect relationship hypotheses with a maximum, user specified depth. Efficiently pruning the non prominent communities and examining the closeness of the ones that can potentially be prominent. They used Named Entity Tagger, Apriori algorithm, open and closed discovery algorithms described in Srinivasan (2004)

Zuraini Zainol et al. in 2017 \cite{zainol2018visualurtext} discusses about the development of text analytics tool that is proficient in extracting, processing, analyzing the unstructured text data and visualizing cleaned text data into multiple forms such as Document Term Matrix (DTM), Frequency Graph, Network Analysis Graph (based on  co-occurrence), Word Cloud and Dendogram (tree structured graph).

Byung-Kwon Park and Il-Yeol Song in 2011 \cite{park2011toward} performed text mining operations and used XML-OLAP, DocCube, Topic Cube to extract structured data from text documents. Based on the schema of structured data. The architecture is based on the concept of OLAP, consolidation OLAP, which integrates two heterogeneous OLAP sources relational OLAP and Text OLAP. Consolidation OLAP enables users to explore structured and unstructured data at a time for getting total business intelligence, and to find out the information that otherwise could be missed by relational or text OLAP when processed separately.

Shane Ahern et al. in 2007 \cite{ahern2007world} developed a sample application that generates aggregate location-based knowledge from unstructured text associated with geographic coordinates. Used the k-Means clustering algorithm, based on the photos’ latitude and longitude. Author used a TF-IDF approach that assigns a higher score to tags that have a larger frequency within a cluster compared to the rest of the area under consideration.

K.L.Sumathy and M.Chidambaram in 2013 \cite{sumathy2013text} gives an overview of the concepts, applications, issues and tools used for text mining. Steps involved in text mining are discussed. A list of tools used is also given.

Paolo Nesi et al. in 2014 \cite{nesi2014ge} used part-of-speech-tagging, pattern recognition and annotation for extracting addresses and geographical coordinates of companies and organizations from their web domains.

Ammar Ismael Kadhim et al. in 2014 \cite{kadhim2014text} proposed processing of unstructured data. Their methodology includes preprocessing (stemming, removing stop words, removing highly frequent and least frequent words). TF-IDF was used for document representation. SVD was used to reduce high dimensional dataset to lower dimension.

\subsection{Question Answer generation techniques}
\label{S:2.2}

Inter AI Developer Program in 2019 \cite{intelnlp2018} developed a system which can generate logical questions from given text input that only humans are capable of doing. Their process involved of selecting important sentence, using parser to extract NP and ADJP from important sentences as candidate gaps and then generate fill in the blanks and brief type questions using NLTK parser and grammar syntax logic. They have succeeded in forming two types of questions fill in the blanks statement and fully stated questions.

Alec Kretch in 2018 \cite{medium2018} developed T2Q for various types of questions using US history dataset. A list of subjects and corresponding phrases was built from the tokens and POS tags using text chunking, Stanford Dependencies and a multiclass classifier. Universal Dependencies were extracted such as from using Stanford CoreNLP’s UDFeatureAnnotator then some words were converted to a Synonym using Princeton WordNet or Glove. T2Q also matches common pattern that identify the date/year of a given event. Questions of each question type for each subject/phrase were formed using basic pattern of each question type as guide. Then similar, but false, answers for each question based on the question type using patterns of good false answers for each question as a guide were formed.

Athira P. M. et al. in 2013 \cite{athira2013architecture} describes the architecture of a Natural Language Question Answering (NLQA) system for a specific domain based on the ontological information, a step towards semantic web question answering. Their main steps include syntactic analysis , semantic analysis, Question Classification, Query Reformulation. They also performed answer filtering and answer ranking. Their results showed that they were able to achieve 94 \% accuracy of natural language question answering in their implementation.

Ming Liu et al. in 2012 \cite{liu2012using} proposed a semi automatic question generation to support academic writing. Key concepts were identified using an unsupervised algorithm to extract key phrases from an academic paper. The system then classifies each key phrase based on a Wikipedia article matched with the key phrase by using a rule-based approach then Wikipedia was used as a domain knowledge base. Knowledge from a single article is used to build conceptual graphs used to generate questions. To evaluate the quality of generated questions a Bystander Turing test was conducted which showed good system rating.

Eiichiro S. et al. in 2005 \cite{sumita2005measuring} proposed a technique of automatic generation of fill in the blanks questions (FBQs) together with testing based on Item Response Theory (IRT) to measure English proficiency. A method based on item information was proposed to estimate the proficiency of the test-taker by using as few items as possible. Results suggest that the generated questions plus IRT estimate the non-native speakers’ English proficiency while on the other hand, the test can be completed almost perfectly by English native speakers. The number of questions can be reduced by using item information in IRT.

Yllias Chali and Tina Baghaee in 2018 \cite{chali-baghaee-2018-automatic} proposed a sequence to sequence model that uses attention and coverage mechanisms for addressing the question generation problem at the sentence level. The attention and coverage mechanisms prevent language generation systems from generating the same word over and over again, and have been shown to improve a system’s output. They have used the simple RNN encoder-decoder architecture with the global attention model. Further, they applied a coverage mechanism, which prevents the word repetition problem. Experimental results on the Amazon question/answer dataset showed an improvement in automatic evaluation metrics as well as human evaluations from the state-of-the art question generation systems.

Xinya Du et al. in 2017 \cite{du2017learning} framed the task of question generation as a sequence-to-sequence learning problem that directly maps a sentence from a text passage to a question. Their approach is totally data driven and requires no manually generated rules. They modeled the conditional probability using RNN encoder-decoder architecture and adopt the global attention mechanism to make the model focus on certain elements of the input when generating each word during decoding. They investigated two variations of their models: one that only encodes the sentence and another that encodes both sentence and paragraph level information. Automatic evaluation results showed that their system significantly outperforms the state of the art rule-based system. In human evaluations, questions generated by this method are also rated as being more natural.

\section{Proposed Methodology}
\label{S:3}

\subsection{Overview of the Work}
\label{S:3.1}

Before describing the detailed outline of this architecture, I shall use a real example to show how exactly it corresponds to the way a human tackles such problems. Imagine a scenario where a student has to find a particular word in a book without an appendix section. To perform this search operation, the most effective way is to know which domain this particular term belongs to and hence a relation of that can be found on the chapter name. After reaching that chapter, the student can relate to which particular section the term may be occurring and hence selects that. If the student is unable to find the term in that section, then he repeats it for the other sections that are related, until the term is found. The other non-human way is maybe searching the term throughout the book either serially or randomly. To solve this type of problem in computers, advanced and interactive techniques must be used to get down to the final result from a huge collection of unstructured data collected on a particular query.

    \begin{figure}[H]
      \begin{center}
        \includegraphics[width=10cm]{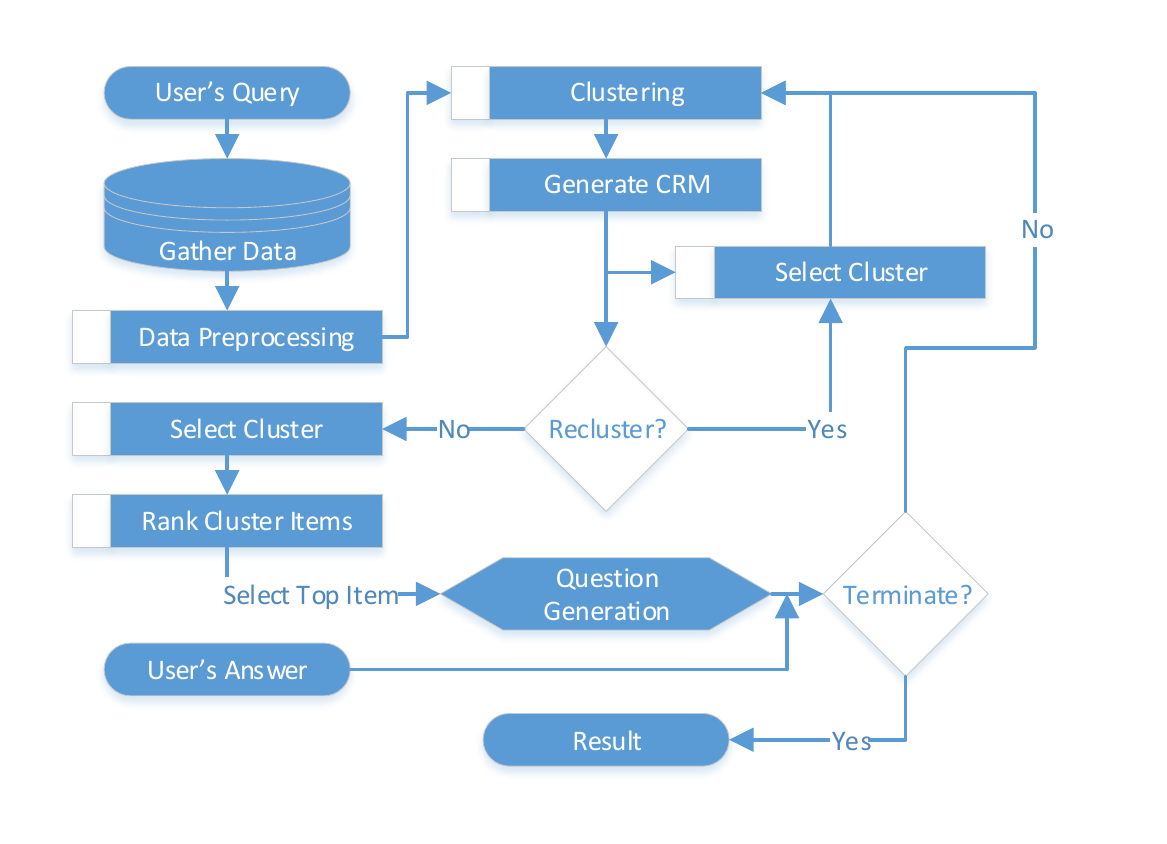}
        \caption{The proposed model framework}
        \label{fig:main_flow}
       \end{center}
    \end{figure}

To carry out the interaction based information retrieval, a model is proposed as shown in Figure-\ref{fig:main_flow}. The two main objectives of this model is to find a word or a sentence from which a question can be developed and a specialized natural language question generator that can generate questions using either a word or a sentence. First phase of this model deals with the objective of identification of questionable entity. To identify a questionable entity, the unstructured raw data can be clustered into various sections and using interaction mechanisms it can be converged into a particular information that the user is looking for. The model consists of a section that retrieves unstructured data from the web following a query provided by the user. That raw data contains many unnecessary tokens that needs to be either removed or changed in order to make the later processes more accurate. The preprocessed tokens(words here) are fed into the clustering section. The K-Means clustering algorithm is selected for clustering as per the experiment in Section-\ref{S:3.2} and research shows K-Means to be beneficial for text data clustering \cite{kadhim2014text}. After clustering is complete, a technique is proposed to identify the relations between the clusters in a semantic way. This technique is named as Cluster Relationship Matrix (CRM). Using the CRM, a semantic distance is calculated and provided with a hyper parameter threshold $\delta$. A mechanism is developed to identify cluster quality and tell whether any further re-clustering is required or not. If re-clustering is required, then a cluster using Algorithm 2 (Step-6) is selected. Similarly, if re-clustering is not required then a cluster is selected according to Algorithm 2 (Step-5) for ranking. A ranking mechanism is used to rank the items present in the selected cluster. After the cluster items are ranked, the sentence containing that particular item is selected and passed to the question generation phase.

The above section describes the process used to identify a questionable entity. Once identified, this entity (basically a word or a sentence from which question can be formed) is passed to a sequence-to-sequence\cite{Sutskever:2014:SSL:2969033.2969173} encoder-decoder RNN model\cite{cho-al-emnlp14} to generate questions. A hierarchical attention model\cite{Yang2016HierarchicalAN} is introduced into the attention layer for this question generation model to provide both word based and sentence based attention discovery. This helps in generation of quality questions both for input word or sentences. Later, analysing the model using the provided answer is used to estimate the expected conclusion of the job and whether to terminate or continue the process. After termination, the final output result is provided to the user.

\subsection{Selection of Clustering Algorithm}
\label{S:3.2}

Clustering is an integral part of machine learning process with unlabelled data. Selection of suitable clustering algorithm plays an important role in making the process of data discovery more accurate and efficient. The data to be clustered in this paper has two important properties, firstly, they are of text and numerical type and secondly they are collected with respect to a particular query. As per the clustering is concerned, the words are selected as tokens to be clustered. For testing the capability of clustering algorithms on text type data relevant to this work, a basic experiment is performed with a text corpus created by combining datasets from three labelled dataset. The clustering algorithms can be broadly divided into three types. So for the experiment, one of the mostly used algorithms in this particular domain is selected from the following three categories.

\begin{enumerate}
    \item \textbf{\textit{Partition based clustering}}- In partition clustering, each and every point in the dataset is assigned to exactly one of the K clusters formed by the algorithm where K is a parameter that has to be passed to the algorithm there are various methods available to find optimal K \cite{kadhim2014text}. The most widely used algorithm that belongs to this category is K-Means. K-Means algorithm scales very well with large data \cite{scikit-learn, scikit:2.3}. K-Means basically try to minimize WCSS (within cluster sum of squares). K-Means with cosine similarity is spherical clustering \cite{banerjee2005clustering}.
    
    \item \textbf{\textit{Hierarchical clustering}} - Hierarchical clustering generates a tree structure of the nested group of clusters\cite{kadhim2014text}. The Agglomerative Clustering is a widely used algorithm in this category. It is a bottom-up algorithm It initially treats each data point as a cluster and then they are successively merged together. For merge strategy, a metric is used known as linkage criterion \cite{scikit-learn, scikit:2.3}.

    \item \textbf{\textit{Density-based clustering}} - The idea behind density-based clustering is to find clusters as high-point density that separates the area of low-point density which leads to an arbitrary shape cluster. High-density points considered to be within clusters and low-density points are considered to be noise. One of the most widely used algorithms in this category is GMM.  GMM assumes there are K Gaussian distributions from which each and every data point was generated. All these K Gaussian distributions have unknown parameters. GMM can be thought of as an extension of K-Means where information about the covariance structure of the data and the centres of the latent Gaussians are also incorporated into the algorithm \cite{scikit:2.1}.

\end{enumerate}
\vspace{0.5cm}
\par For clustering evaluation, four algorithms are selected namely Sphere clustering, KMeans with Euclidean, Agglomerative clustering and Gaussian Mixture Model Clustering (GMM) due to their high accuracy in text data. The steps associated with the selection of the best algorithm for this particular experiment are preprocessing, embedding and finally the selection, are discussed in the sub-sections below.

\subsubsection{Preprocessing}
\label{S:3.2.1}
The web sources used to get labelled data had data arranged in tabular, list format, so not much preprocessing was required. Therefore, only basic preprocessing was done before using Sentence transformers for embedding using the steps discussed below.

    \begin{algorithm}[H]
    \floatname{algorithm}{Algorithm 1}
    \renewcommand{\thealgorithm}{}
    \caption{Preprocessing for clustering algorithm selection stage}
    \label{algo:preprocessing}
    \begin{algorithmic}[1]
    \STATE Replaced “\_” with spaces
    \STATE Converted all characters to lowercase
    \STATE Replaced “TABS” with white spaces
    \STATE Removed all special characters
    \STATE Removed all multiple spaces
    \STATE Applied lemmatisation
    \end{algorithmic}
    \end{algorithm} 

After preprocessing is complete, three lists are generated. One containing symptoms with 312 number of items, one containing diseases with 315 number of total items and the other containing medicines with a total of 300 items. 

\subsubsection{Embedding}
\label{S:3.2.2}
Embedding is a form of representation of text data in a multidimensional space, where their semantic meaning and relation is conserved. Embedding are very important in training machine learning models as they provide the source of knowledge of the unknown data to the model. An embedding can be trained with our own data or a pre-trained model can be used. In this paper a well known pre-trained embedding model Sentence Transformer embeddings with “distilroberta-base-paraphrase-v1” model is used \cite{DBLP:journals/corr/abs-1908-10084}. This gives 768 dimensional embedding for the data. Sentence Transformers were used because most of the symptoms, medicines or even diseases are not single word, instead they are combination of multiple words. The distilroberta-base-paraphrase-v1 model is trained on millions of paraphrase examples. They create extremely good results for various similarity and retrieval tasks. Further, PCA was used to reduce dimension of data from 768 to 200 as High dimensional data is not suitable for clustering. 

\subsubsection{Selection}
\label{S:3.2.3}

Four clustering algorithms were selected as discussed, namely KMeans, Sphere clustering (KMeans with cosine similarity), Agglomerative clustering and GMM clustering. The experiment for selection of the algorithms used various parameters and considerations that are provided below.
\begin{itemize}
    \item \textbf{\textit{K-Means -}} K-Means was used with the number of clusters as 3 and with ‘K-Means++’ as method of initialization to speed up the convergence.
    \item \textbf{\textit{Spherical clustering -}} All parameters used were the same as K-Means but this uses cosine similarity whereas K-Means uses euclidean similarity.
    \item \textbf{\textit{Agglomerative clustering -}} The number of clusters were taken as 3. ward was chosen as linkage criterion which minimizes the variance of the clusters being merged. Euclidean metric was used to compute the linkage.
    \item \textbf{\textit{GMM clustering -}} The number of mixture components were taken as 3 with co-variance type as full which means each component has its own general co-variance matrix.
\end{itemize}

\subsection{Identification of Questionable Entity}
\label{S:3.3}

    \subsubsection{Multilayered Clustering}
    \label{S:3.3.1}
    An algorithm is proposed to retrieve useful information from an unstructured preprocessed text corpus. The input to this algorithm is unstructured text and the objective output is a list of ranked (ranking is defined in subsequent sections) words from the cluster having maximum avg distance to all other clusters. Here, the distance is calculated via a novel cluster distance metric developed specifically for the type of the data, this paper is dealing with. The algorithm for multilayered clustering is provided below:
    
    \begin{algorithm}[H]
    \floatname{algorithm}{Algorithm 2}
    \renewcommand{\thealgorithm}{}
    \caption{Multilayered Clustering}
    \label{algo:multilayered_clustering}
    \begin{algorithmic}[1]
    \STATE First, the given text corpus is clustered into N (= 3) initial clusters using KMeans. We have chosen N=3 initially because our main focus on our study is on medical text corpus which generally consists of three group of words, which are medicines/treatments, diseases \& symptoms.
    \STATE Then a cluster relationship matrix(CRM) is created of size NXN where CRM[i][j] = distance metric of cluster pair i, j. This distance metric indicates how distinct two clusters are with respect to each other. Higher the distance metric farther the two clusters are. Detailed description of the CRM is provided in Section-\ref{S:3.3.2}.
    \STATE Then an un-directed graph is created with each node as a cluster (0 to N-1) and M[i][j] as the weight of an edge connecting two nodes(clusters) $i$, $j$. Now for each node sum of weights of all edges directly connected to it is calculated (there exist N-1 such edges for each node). Now out of all N nodes, two nodes are selected for which this calculated value is maximum and minimum let’s call these nodes as $maxDisNode$ and $minDisNode$. The clusters corresponding to these two nodes are chosen as $maxDisCluster$ and $minDisCluster$.
    \STATE Then carefully a hyper-parameter threshold '$\delta$' is chosen (this threshold can be dynamically updated and tuned with each iteration of the algorithm). Now if each and every value of M[i][j] in $N\times{N}$ cluster relationship matrix is less than the threshold then reclustering (Step 6) is required else ranking (Step 5) is selected.
    \STATE Ranking - In this phase, $maxDisCluster$ is picked and then algorithm sort the words in that cluster according to the increasing euclidean distance that the vector representation of that word has from the centre of the cluster and then the algorithm ends.
    \STATE Reclustering - In the reclustering $minDisCluster$ is picked and then algorithm divides that cluster into two clusters so the total number of the cluster now changes from N to N+1. Accordingly, the dimension of the cluster relationship matrix changes to (N+1)X(N+1) and its values also update accordingly. Now go to step 4.
    \end{algorithmic}
    \end{algorithm} 
    
    \subsubsection{Cluster Relationship Matrix}
    \label{S:3.3.2}
    Our proposed algorithm takes $N$ clusters as an input and gives cluster relationship matrix as the output. It is an $N\times{N}$ matrix where $N$ is the total number of clusters present at the given moment where $M[i][j]$ = distance metric of cluster pair i, j. This distance metric indicates how distinct two clusters are with respect to each other. Higher the distance metric more distinct the two clusters are (varies between zero to one). To calculate $M[i][j]$ (distance metric of cluster pair i, j) a special distance Word Movers Distance (WMD) is calculated between each pair of words from cluster $i$ and cluster $j$, and then are summed up. Finally, the whole matrix is normalized to have values between 0 and 1.
    
    \paragraph{Word Movers Distance (WMD):}
    \label{S:3.3.2.1}
    Term frequency-inverse document frequency (TF-IDF) and the bag of words (BOW) are two most common ways to represent documents however, these ways are frequent near orthogonal so these are not suitable for document distances. Also, the distance between individual words is not captured in these ways. Many attempts have been made to solve this problem most of which learned a latent low dimensional representation of documents. Although these attempts produced a more reasonable representation of documents as compared to BOW but they in many cases do not improve distance-based tasks performance of BOW. Due to these limitations WMD was introduced to capture distance between the two documents in a much better way. In our case, we are trying to keep words with less distance together in the same cluster WMD metric is suitable for our use case. WMD performance was evaluated using KNN on 8 document datasets and compared it with BOW, TFIDF, BM25 LSI, LDA, mSDA, CCG. The results were impressive. It was found out that WMD performed best on six out of eight datasets. Even on the two datasets on which WMD was not the best performer the error rates were very close to the top performers.
    
    WMD uses word embeddings like word2vec, GloVes, etc that learns semantic meaningful vector representation of words from their local co-occurrences in the sentence. MWD measures the semantic distance between two documents i.e. it measures the minimum amount of distance that the embedded words of one document need to “travel” to reach the embedded words of another document. It does not require any hyperparameters. It is highly interpretable and has high retrieval accuracy. The WMD distance between two documents is defined by the three main parts as described below:
    \begin{enumerate}
        \item \textbf{\textit{Document representation}} - The document is represented as an $n$ dimensional $d \in R^n$ vector where $n$ is the vocabulary size. Each element of this vector is a word’s normalized frequency in the document. This vector is also known as normalised bag of words (nBOW) vector. If a word i appears $C_i$ times in the document then $d_i$ is found using equation-\ref{eq:di}. This vector is usually very sparse.
        
        \begin{equation}
            d_{i} = \frac{C_i}{\sum_{j=1}^{n}C_j}
            \label{eq:di}
        \end{equation}

        \item \textbf{\textit{Semantic similarity metric}} - The travel cost$C(i, j)$ from word $i$ in one document to word $j$ in another document is defined in equation-\ref{eq:cij}.
        
        \begin{equation}
            C(i,j) = \left | \left | X_i - X_j \right | \right |_2
            \label{eq:cij}
        \end{equation}
        where, $X_i$and $X_j$ are embedding representation of words $i$ and $j$ respectively.
        
        \item \textbf{\textit{Flow matrix}} - The flow matrix $T \in R^{n\times{n}}$ where, $n$ is the vocabulary size is a sparse matrix where $T_{ij} \geq 0$ denotes how much of word $i$ in one document travels to word $j$ in another document
    \end{enumerate}

    \paragraph{Ranking}
    \label{S:3.3.2.2}
    This algorithm takes $maxDisCluster$ as an input and gives a ranked list of words from that cluster as an output. The ranking is an effort to find the most representative words of a cluster. The words with a higher ranking are better representative of the cluster as compared to words with a lower ranking. Once no reclustering is required the algorithm selects $maxDisCluster$ (defined in Section-\ref{S:3.3.1}) and sort words (words appearing first have higher rank) in it. Sorting is done using Algorithm-3.
    
    \begin{equation}
    D_i = \sqrt{\sum_{j=1}^{K} (C_j - E_{ij})^2}
    \label{eq:Di}
    \end{equation}
    where, $K$ is the dimension of the embedding used.
    
    \begin{algorithm}[H]
    \floatname{algorithm}{Algorithm 3}
    \renewcommand{\thealgorithm}{}
    \caption{Semantic ranking of elements in the selected cluster}
    \label{algo:ranking_list}
    \begin{algorithmic}[1]
    \STATE Let the centre of the $maxDisCluster$ be $C$.
    \STATE Let our cluster has $M$ words in it and let the $i^{th}$ word be $W_i$ and its embedding representation be $E_i$
    \STATE For each $i^{th}$ word $W_i$ in $maxDisCluster$ euclidean distance between $E_i$ and $C$ is calculated let this calculated value be $D_i$ as shown in equation-\ref{eq:Di}.
    \STATE Now, all $M$ words in $maxDisCluster$ are sorted based on distance $D_i$ i.e. the words having smallest $D_i$ will come first.
    \STATE The final sorted list is returned.
    \end{algorithmic}
    \end{algorithm} 
 
\subsection{Natural Language Question Generation}
\label{S:3.4}

    \subsubsection{Overview}
    \label{S:3.4.1}
    Natural language question generation is an important aspect of Natural Language Processing field in computer science due to recent advancement in education systems and interaction based autonomous systems. This section of the paper focuses on the methodology followed in generation of questions relevant to medical domain when provided either with a word or a sentence$D={W_n}$, where $n-1$ is the total number of words present in the sentence and $n=0$ represents a single word. Here, the main objective is to maximize accuracy of the conditional probabilities P($Q|W_n$); where $Q$ is the question and $W_n$ is the input provided to the system of which a question needs to be generated (Equation-\ref{eq:obj}).
    
    \begin{equation}
        p(Q|W_{n}) = \prod_{1}^{n}p(q_{t}|q_{1:t-1}, W_{n})
        \label{eq:obj}
    \end{equation}
    
    For generation of natural language question, a Sequence-to-Sequence learning \cite{Sutskever:2014:SSL:2969033.2969173} Encoder Decoder model for Recurrent Neural Networks (RNN) \cite{Sutskever:2014:SSL:2969033.2969173} is used along with improved coverage mechanism\cite{chali-baghaee-2018-automatic} for generating natural language questions for the input data $D$. A generic attention layer provides attention discovery for multiple words and usually suffer when sentences consists of lower number of words or if it a single word. To overcome this issue, Hierarchical Attention Network (HAN)\cite{Yang2016HierarchicalAN} is used along with the general attention layer; this combined attention model ensures the classification from both word level and sentence level. So that it can identify the presence of words that convey a higher weight to the context of the overall sentence. Since, HAN doesn't consider long term information about the previous sentences, it causes to loose information for longer sentences. Hence, the combined attention model can select the highest attention weights whenever applicable, in order to keep the context of the sentence through longer sentences. A well known pre-trained embedding GloVes is used and this model is trained, validated, and tested by both the AmazonQA dataset and PubMed datset.

    \subsubsection{Hierarchical Attention Network}
    \label{S:3.4.2}
    
    The content provided as training, contains sample of both, more and less important data. Knowledge of this content is usually achieved using attention mechanisms over the sequence of data. But the attention models usually face difficulty to identify the contribution of words for smaller sequence of inputs. To overcome the issue of attention discovery for smaller sequences of inputs and even for words, a hierarchical attention network(HAN) \cite{Yang2016HierarchicalAN} proves to be a good option. It has two levels of attention mechanisms applied at the word and sentence-level, enabling it to attend differently to more and less important content. HAN calculates the attention in terms of words and sentences hence two context vectors are used, one for word level attention and another for sentence level attention. The process HAN starts by word level encoder followed by a attention mechanism only for words (Equation-\ref{eq:attn_han_words}), then those attention vectors are passed to the sentence level encoders followed by the sentence attention (Equation-\ref{eq:attn_gen}).
    
    \begin{equation}
    \label{eq:attn_1}
        u_{it} = tanh(W_{w}h_{it}+b_{w})
    \end{equation}
    Where, $W_{s}$ are the weights associated with the concatenated hidden states $h_{it}$ and $b_{w}$ is the bias.
    
    \begin{equation}
    \label{eq:attn_2}
        \alpha_{it} = \frac{exp(u_{it}^{T}u_{w})}{\sum_{t}exp(u_{it}^{T}u_{w})}
    \end{equation}
    Where, $u_{it}$ is the hidden representation of $h_{it}$ and $u_{w}$ is word level context vector. $T$ is the total number of input sequences. The value of $u_{it}$ is calculated from Equation-\ref{eq:attn_1}.
    
    \begin{equation}
    \label{eq:attn_han_words}
        s_{i} = \sum_{t}\alpha_{it}h_{it}
    \end{equation}
    
    Where, $s_{i}$ denotes the word level attention, and $\alpha_{it}$ is calculated from Equation-\ref{eq:attn_2} and $h_{it}$ are the concatenated hidden states of the network for a bi-directional model. For sentence level attention, the general attention model (Equation-\ref{eq:attn_gen}) is selected.
    
    \subsubsection{Encoder Modelling}
    \label{S:3.4.3}
    The input and output in a question generation model is dynamic in nature due to the fact that the input can have varying number of words and same for the output question. To deal with these kind of inputs and to convert the variable length input data into a fixed length output, an encoder is used. It is a set of neural network layers that maps the input data $D$ into word vectors, which are populated in the network as hidden states $H = \{h_1,h_2,...,h_n\}$, where $D={W_n}$. For this experiment, the selected encoder is a bidirectional LSTM layer \cite{hochreiter1997long}. Two layers of the bidirectional LSTM is considered in this experiment, as it is already proven to be good in similar context \cite{chali-baghaee-2018-automatic}. The bidirectional LSTM layer has two hidden states, one is a forward state $f_i$ and the other is a backward state $b_i$, for an index $i$, enabling the network to understand a sentence and its formation from both starting and from the ending. Considering a input index of $i$, the hidden states are concatenated to form a longer sequence of hidden states such that $h_i = \{f_i, b_i\}$. The sequence of hidden states $h_i$ is used during decoding for generation of output vector $q_t$ as a method to identify the source and predicting the next target word. Target word in question, $q_t$ is calculated as the weighted sum of hidden states $h_i$ as expressed in Equation-\ref{eq:target_qt}.
    
    \begin{equation}
        \label{eq:target_qt}
        q_{t} = max[\sum_{i = 1}^{n} a_{t}(n)H(n), \sum_{i = 1}^{n} s_{i}(n)H(n)]
    \end{equation}
    
        where, $a_{t}$ and $s_{i}$ are shift vectors and are calculated according to the general attention model and hierarchical attention model (described in Section-\ref{S:3.4.2}) respectively. 
        
    \begin{equation}
        \label{eq:attn_gen}
         a_{t}(i) = \frac{exp(h_{t}^{T}W_{a}h_{i})}{\sum_{j}exp(h_{t}^{T}W_{a}h_{i})}
    \end{equation}
     
    \subsubsection{Decoder Modelling}
    \label{S:3.4.4}
    A sequence-to-sequence model requires a decoder in order to convert the fixed length context output provided by the encoder and convert it to a variable length output. Two layers of LSTM is used as a decoder in the experiment. To avoid attending repetitive words, the attention model provides coverage vector for each input in the time step. The attention model must attain to the next input taking into consideration the previous inputs using the coverage vector $c$ at a time step $t$. 
    
    \begin{equation}
        c_t = \sum_{{t}'=0}^{t-1}a{t}'
    \end{equation}

    Further, the coverage vector needs to be integrated with the concatenated hidden states, and this is done using a $'tanh'$ operation over the hidden states $h_i$ and point wise addition of coverage vector $c_t$ and the weights of the coverage vector $w_c$ to be learned.
    
    \begin{equation}
        h_i = tanh(h_i + w_c c_t (i))
    \end{equation}
    
    The context vector $s_t$ and the output of the previous layers $[z_1,z_2, ..., z_{(t-1)}]$ in the decoder are fed to the next layer in the current time step to make the prediction. The prediction is made using a fully connected layer with a softmax classifier. The attention hidden state $\tilde{h_t}$ is calculated as a $'tanh'$ activation operation on the weights of the context vector $w_x$ to be learned, the context vector from the source $s$ and the next target state $q_{t}$. 
    
    \begin{equation}
        \tilde{h_t} = tanh (W_{x} [s_{t};q_{t}])
    \end{equation}

    Finally, the next hidden state is formed using a LSTM cell with the previous hidden state $h_{t-1}$ and the output from the previous state $z_{t-1}$ as an input.
    
    \begin{equation}
        h_{t} = LSTM(z_{t-1}, h_{t-1})
        \label{eq:lstm}
    \end{equation}

    \subsubsection{Training}
    \label{S:3.4.5}
    
    In the training phase, the generated model is trained using the dataset corpus containing sample questions and their respective answers. Considering the total number of training data as $T$, the corpus data is defined as $C_d = (a_i, q_i)_1 ^T$. The model is trained by providing the answers as an input to the model. For a particular sample answer, the model predicts the questions, which is then optimised by minimizing the negative log-likelihood of the training corpus (Equation-\ref{eq:train_sent}). For word level training, the same model is retrained taking each of the single word in the sample answer (not considering common words) (Equation-\ref{eq:train_sent}). 
    
    \begin{equation}
        M_t = \sum_{i=1}^{T} - log p(q_i|a_i)
        \label{eq:train_sent}
    \end{equation}
    where, $M_t$ is the training model and the rest carry their usual meaning.
    
    \begin{equation}
        M_t = \sum_{i=1}^{T} \sum_{j=1}^{G_i} - log p(q_i|a_{ij})
        \label{eq:train_word}
    \end{equation}
    where, $M_t$ is the training model and $G_i$ is the maximum number of words considered as an input for $i^{th}$ corpus answer.
\vspace{0.5cm}
\par To generate the natural language questions from the model, the prediction vector mapping is performed. Due to limited embedding corpus, many corpus words will be new to the model, these unknown tokens are substituted by average attention weight of the words from the source sentence.

\subsection{Elimination and Termination}
\label{S:3.5}

In this phase, after the generation of question, the user is required to provide some answer to the question. The answer to the question needs to be evaluated by the system such that the information and context related to the question is either kept or eliminated. This phase is also responsible regarding continuation or termination of the current session of information retrieval. To perform these operations, the following algorithm is designed.

\begin{algorithm}[H]
    \floatname{algorithm}{Algorithm 4}
    \renewcommand{\thealgorithm}{}
    \caption{Elimination and Termination}
    \label{algo:des_ele}
    \begin{algorithmic}[1]
    \STATE First, the user input $A_u$ (answer) is preprocessed to remove unnecessary and irrelevant tokens.
    \STATE Then, a sentence concatenation operation of the answer and the question is performed by removing the question word like what, when, where, etc. to get a modified $A_u$. 
    \STATE Then, a similarity measurement is performed $WMD(a_i,A_u)$ for $i^{th}$ prediction. 
    \STATE Considering the same hyper-parameter threshold $'\delta'$ as defined in Section-\ref{S:3.3.1}, if $WMD(a_i,A_u) \geq \delta$ then go to step-5 else go to step-6.
    \STATE Corpus data $a_i$ is kept intact and continuation of the steps is performed with the selected list and rest of the clusters are removed. Go to step-6.
    \STATE Corpus data $a_i$ is removed and perform step $Reclustering$ (Algorithm-2) eliminating the row and column corresponding to the removed cluster. If re-clustering is required then continuation of the steps is performed with the rest of the clusters removing the CRM values (refer to Section-\ref{S:3.3.1}). Else the process is terminated.
    \end{algorithmic}
    \end{algorithm} 

\section{Experiments and Results}
\label{S:4}

\subsection{Objective of the Experiment}
\label{S:4.1}
This paper has various experiment associated with the different sections of the work. Hence, after providing a brief info about the experimental environment and the dataset description, the result section is divided into few categories to improve the readability:
\begin{enumerate}
    \item \textbf{Selection of clustering algorithm:} Clustering algorithm plays a vital role in entity selection phase of the work. To select a particular clustering algorithm in the context of our work, a separate experiment is performed to evaluate the performance of various clustering algorithms.
    \item \textbf{Results for Question Generation: } Results associated with the generation of medical based question generation is presented in this section. Expert opinion is also taken into consideration for question quality evaluation.
    \item \textbf{Comparison with the Existing Approaches:} Performance comparison of the proposed system with the existing works using same data are performed. 
\end{enumerate}

\subsection{Experimental Environment}
\label{S:4.2}
    The specification of the system used is as follows:
        O.S. - Windows 10 Professional, 
        CPU - AMD® Ryzen™ 7-3700X Processor, 
        RAM - 32GB DDR4, 
        GPU - NVIDIA GeForce® GTX 1080 Ti. The windows version of the Python-64Bit \cite{10.5555/1593511} with editor IPython notebook \cite{PER-GRA:2007} throughout the experiment. Important modules used in the experiment include tensorflow \cite{tensorflow2015-whitepaper}, packages from NumPy \cite{oliphant2006guide}, SciPy \cite{2020SciPy-NMeth}, scikit-learn \cite{scikit-learn} and Matplotlib \cite{Hunter:2007}.

\subsection{Dataset Description}
\label{S:4.3}

Two different dataset are used in this section, namely Pubmed 200K dataset \cite{dernoncourt-lee-2017-pubmed} and Amazon QA dataset \cite{McAuley:2016:ACS:2872427.2883044, DBLP:journals/corr/abs-1809-00934}; to train the Recurrent Neural Network models at two  different level. Pub-med has a very rich repository of medical data and research articles; in contrast, Amazon QA dataset has a good collection of opinion question and answer data for the actual natural language question formation phase of the experiment. To test the model, the labelled AmazonQA dataset is used; After training and testing the model with varying parameters, a model is selected which provided the highest accuracy using K-fold cross validation technique. After that, this model is fed with the output word or sentence from the proposed method of identification of questionable entity as discussed in Section-\ref{S:3.3}, in turn, the model is capable of providing some relevant questions from the sentences of the article. 

 \begin{table}[!ht]
 \centering
 \begin{tabular}{||c c c c||} 
  \hline
  Dataset & Training & Validation & Testing \\ [0.5ex]  \hline\hline
  Pubmed & 184472 & 46654 & 34195 \\ 
  Amazon QA & 215325 & 65487 & 45574 \\ [1ex] 
  \hline
 \end{tabular}
 \caption{Dataset Statistics }
 \label{table:1}
 \end{table}

\subsection{Results for selection of clustering algorithm}
\label{S:4.4}

\begin{figure}[H]
     \centering
     \begin{subfigure}[b]{0.24\textwidth}
         \centering
         \includegraphics[width=\textwidth]{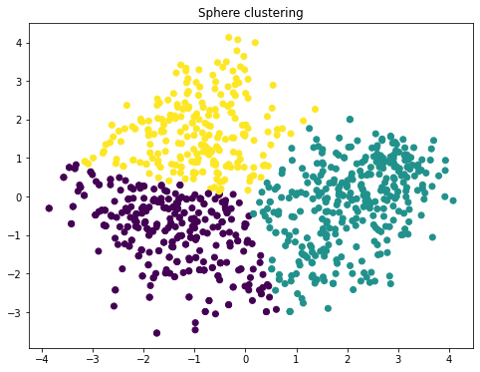}
         \caption{Spehere Clustering}
         \label{fig:ngs}
     \end{subfigure}
     \hfill
     \begin{subfigure}[b]{0.24\textwidth}
         \centering
         \includegraphics[width=\textwidth]{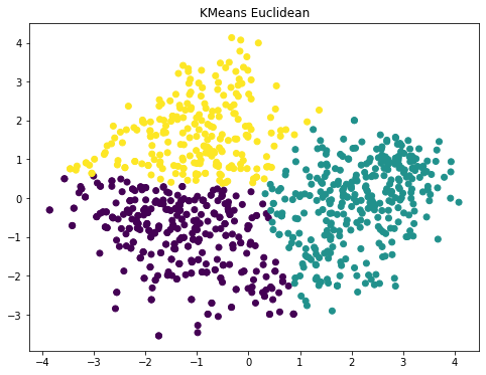}
         \caption{KMeans Clustering}
         \label{fig:nct}
     \end{subfigure}
     \hfill
     \begin{subfigure}[b]{0.24\textwidth}
         \centering
         \includegraphics[width=\textwidth]{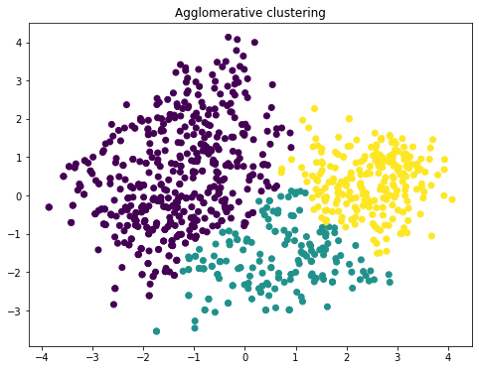}
         \caption{Agglomerative}
         \label{fig:nse}
     \end{subfigure}
     \hfill
     \begin{subfigure}[b]{0.24\textwidth}
         \centering
         \includegraphics[width=\textwidth]{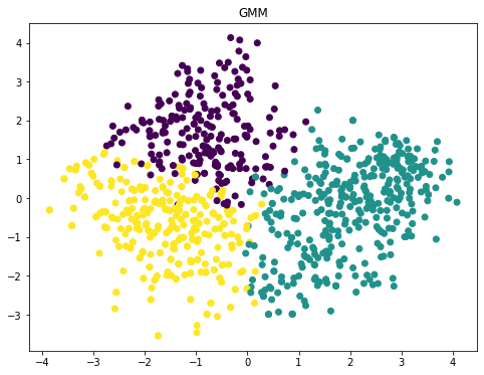}
         \caption{GMM}
         \label{fig:nsd}
     \end{subfigure}
        \caption{Clustering Visualisation}
        \label{fig:cluster_visualize}
\end{figure}

Four top performing clustering algorithms for text related data are selected from each category of the algorithms present. Then the testing has been performed, to select a particular algorithm suitable for this work. The visualization of each of the clustering algorithms are shown in Fig-\ref{fig:cluster_visualize}. From the visualization, it is seen that all of the clustering algorithms are able to detect the clusters with minor variations.  

Bcubed precision, recall, f score was used for evaluation. Bcubed metric is widely used as standard metric for text clustering problems. It satisfy four formal constraints Cluster Homogeneity, Cluster Completeness, Rag Bag and Clusters size vs. quantity. The average BCubed precision is average precision of all items in the distribution similarly the average BCubed recall is average recall of all items in the distribution. The BCubed precision of an item is ratio of number of items present in its cluster that have same category as its (including itself) to the total number of items in its cluster. The BCubed recall of an item is the ratio of number of items present in its cluster that have same category as its (including itself) to the total number of items in its category. For performance evaluation, the general precision, recall and F-score metrics used are provided in the equations \ref{EquationP}, \ref{EquationR}, \ref{EquationF}.

\begin{equation}
    Precision(P) = \frac{TP}{TP+FP}
    \label{EquationP}
\end{equation}

\begin{equation}
    Recall(R) = \frac{TP}{TP+FN}
    \label{EquationR}
\end{equation}

\begin{equation}
    F1-Score(F1) = 2\times \frac{P \times R}{P+R}
    \label{EquationF}
\end{equation}

\begin{table}[H]
\centering
\resizebox{\textwidth}{!}{%
\begin{tabular}{|c|c|c|c|}
\hline
\textbf{Algorithm}                                           & \textbf{Precision} & \textbf{Recall} & \textbf{F score} \\ \hline
\textit{Sphere clustering (KMeans with cosine   similarity)} & 0.71               & 0.71            & 0.71             \\ \hline
\textit{KMeans with Euclidean}                               & 0.74               & 0.75            & 0.75             \\ \hline
\textit{Agglomerative Clustering}                            & 0.68               & 0.70            & 0.69             \\ \hline
\textit{GMM Clustering}                                      & 0.71               & 0.72            & 0.71             \\ \hline
\end{tabular}%
}

 \caption{Evaluation of clustering algorithm on medical texts }
 \label{tab:cluster_eval}
\end{table}

The evaluation result as shown in the Table-\ref{tab:cluster_eval} for the clustering algorithm shows K-Means to be the best performing model and hence is selected for the later stages of the model.

\subsection{Results for Question Generation}
\label{S:4.5}

\subsubsection{Model training and testing performance}
\label{S:4.5.1}

\begin{figure}[H]
     \centering
     \begin{subfigure}[b]{0.45\textwidth}
         \centering
         \includegraphics[width=\textwidth]{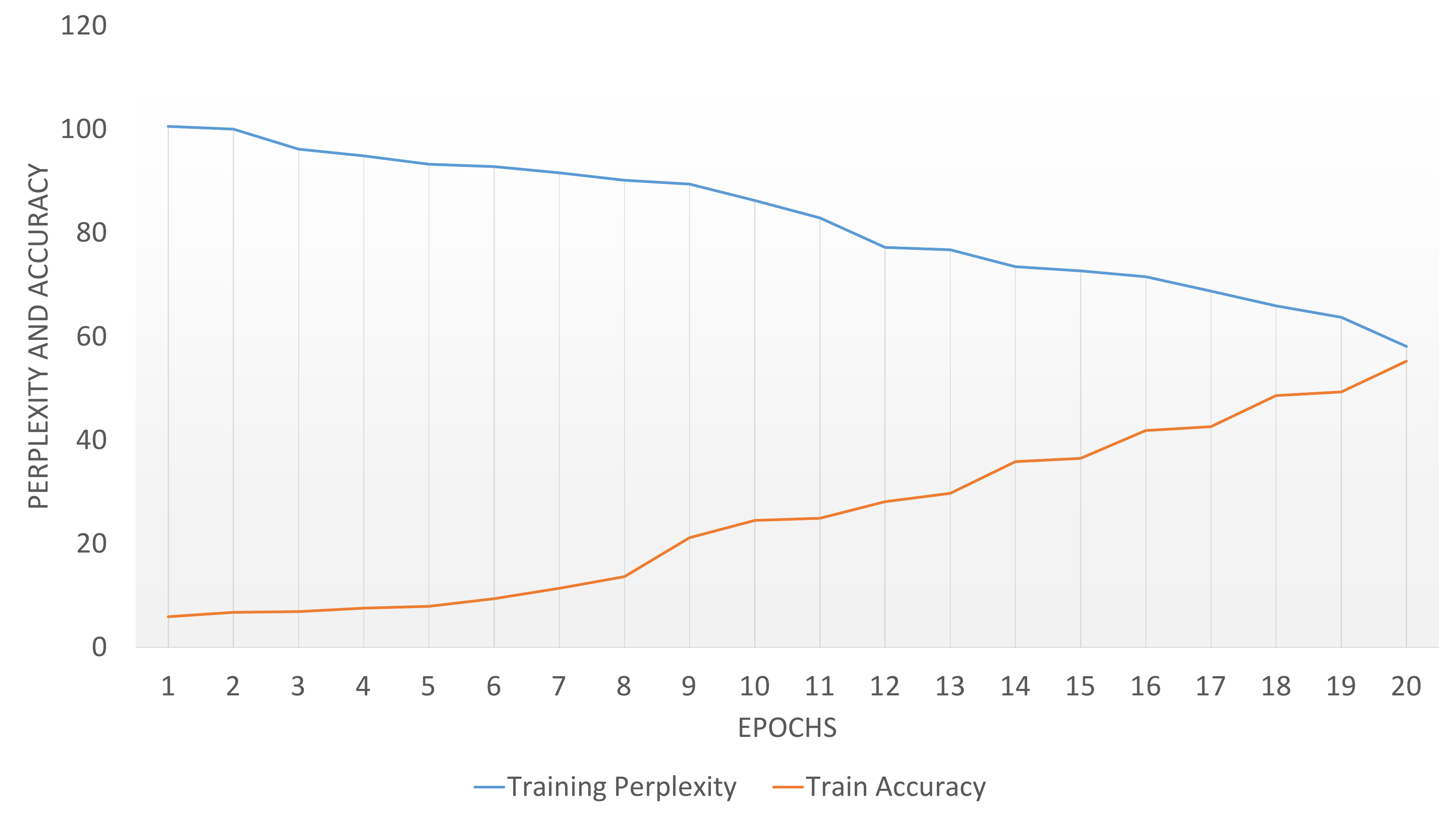}
         \caption{Training Statistics}
         \label{fig:wts}
     \end{subfigure}
     \begin{subfigure}[b]{0.45\textwidth}
         \centering
         \includegraphics[width=\textwidth]{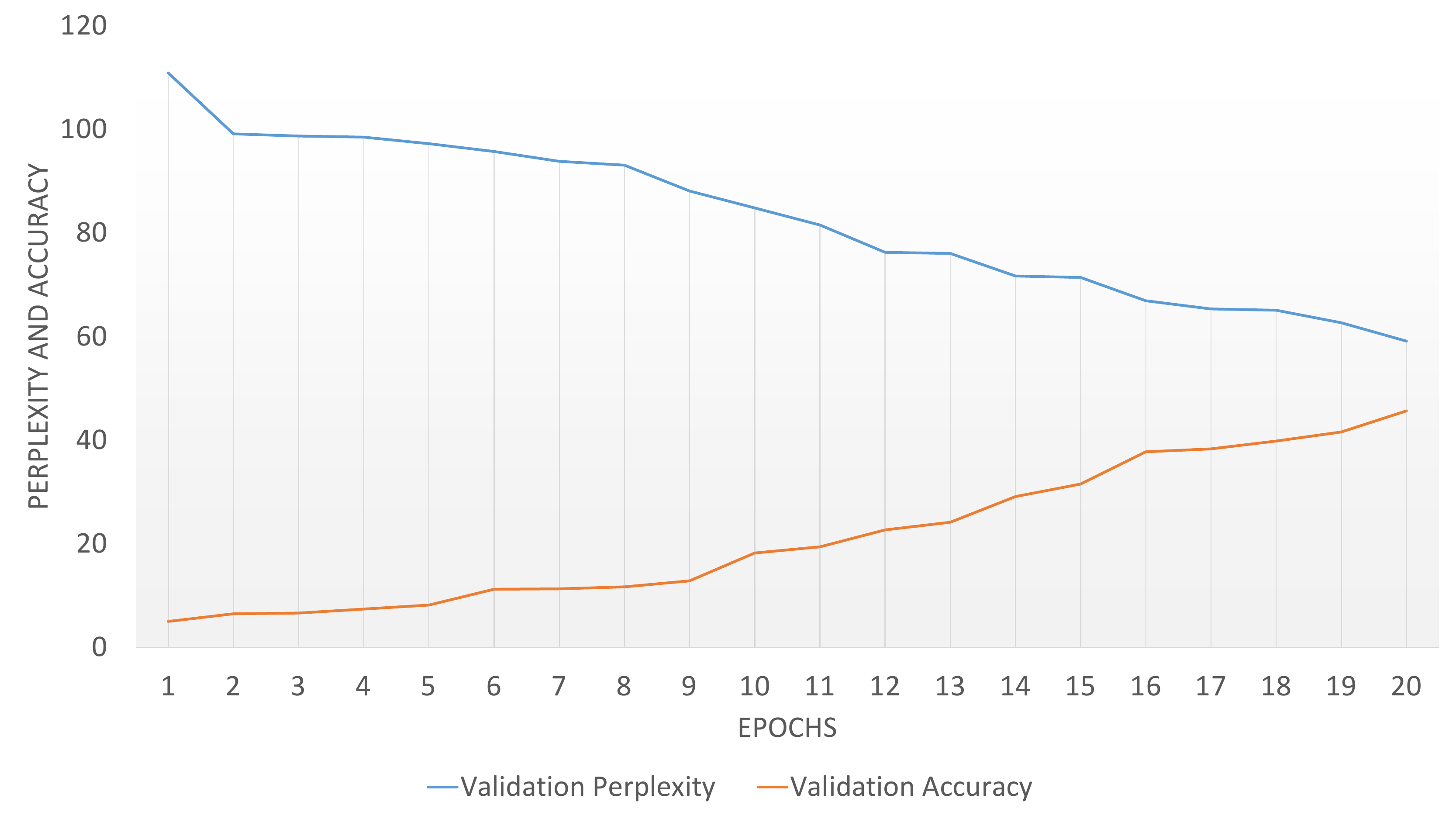}
         \caption{Validation Statistics}
         \label{fig:wvs}
     \end{subfigure}
        \caption{Model perplexity vs accuracy statistics for single words}
        \label{fig:word_train_val}
\end{figure}

\begin{figure}[H]
     \centering
     \begin{subfigure}[b]{0.45\textwidth}
         \centering
         \includegraphics[width=\textwidth]{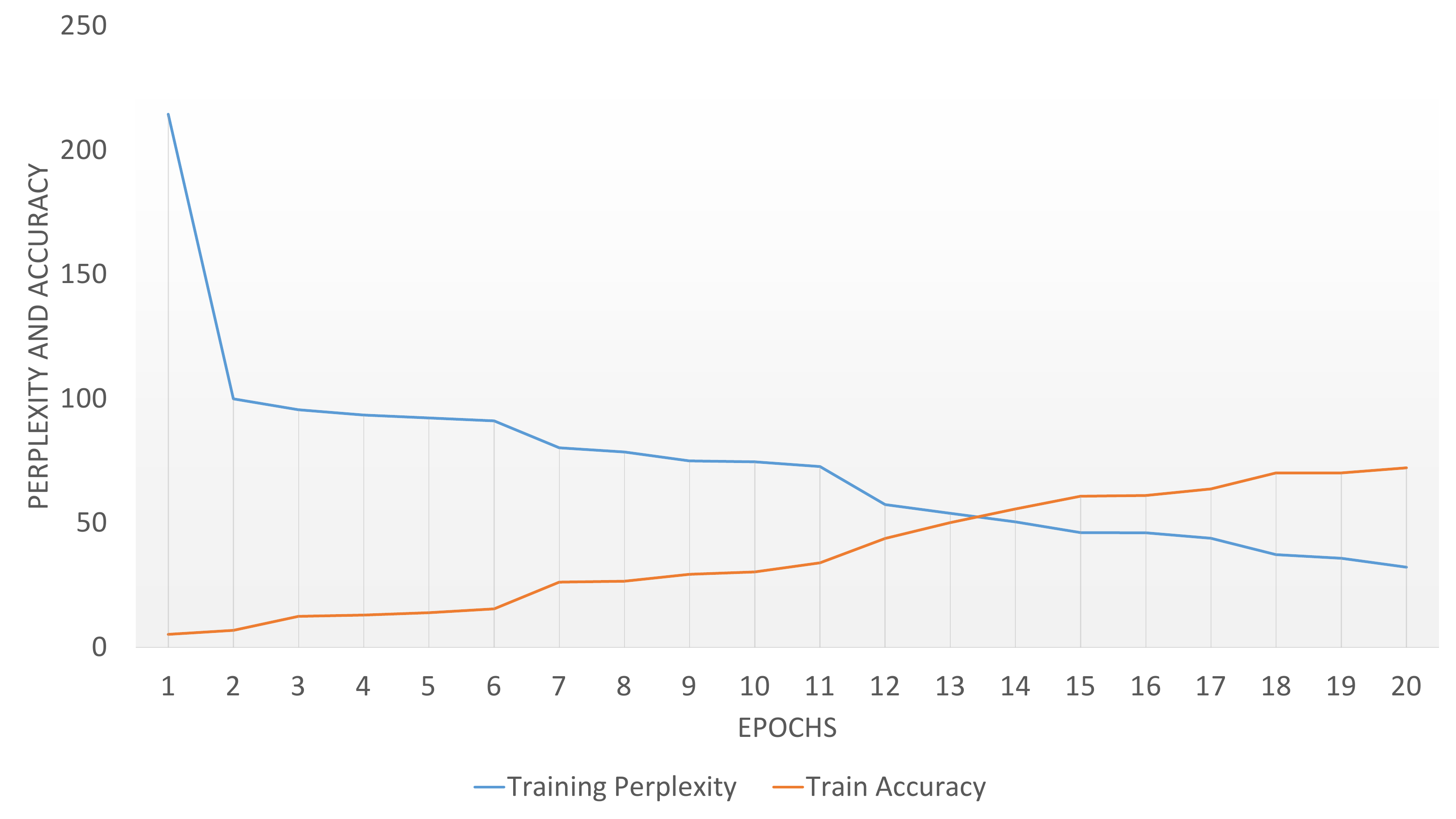}
         \caption{Training Statistics}
         \label{fig:sts}
     \end{subfigure}
     \begin{subfigure}[b]{0.45\textwidth}
         \centering
         \includegraphics[width=\textwidth]{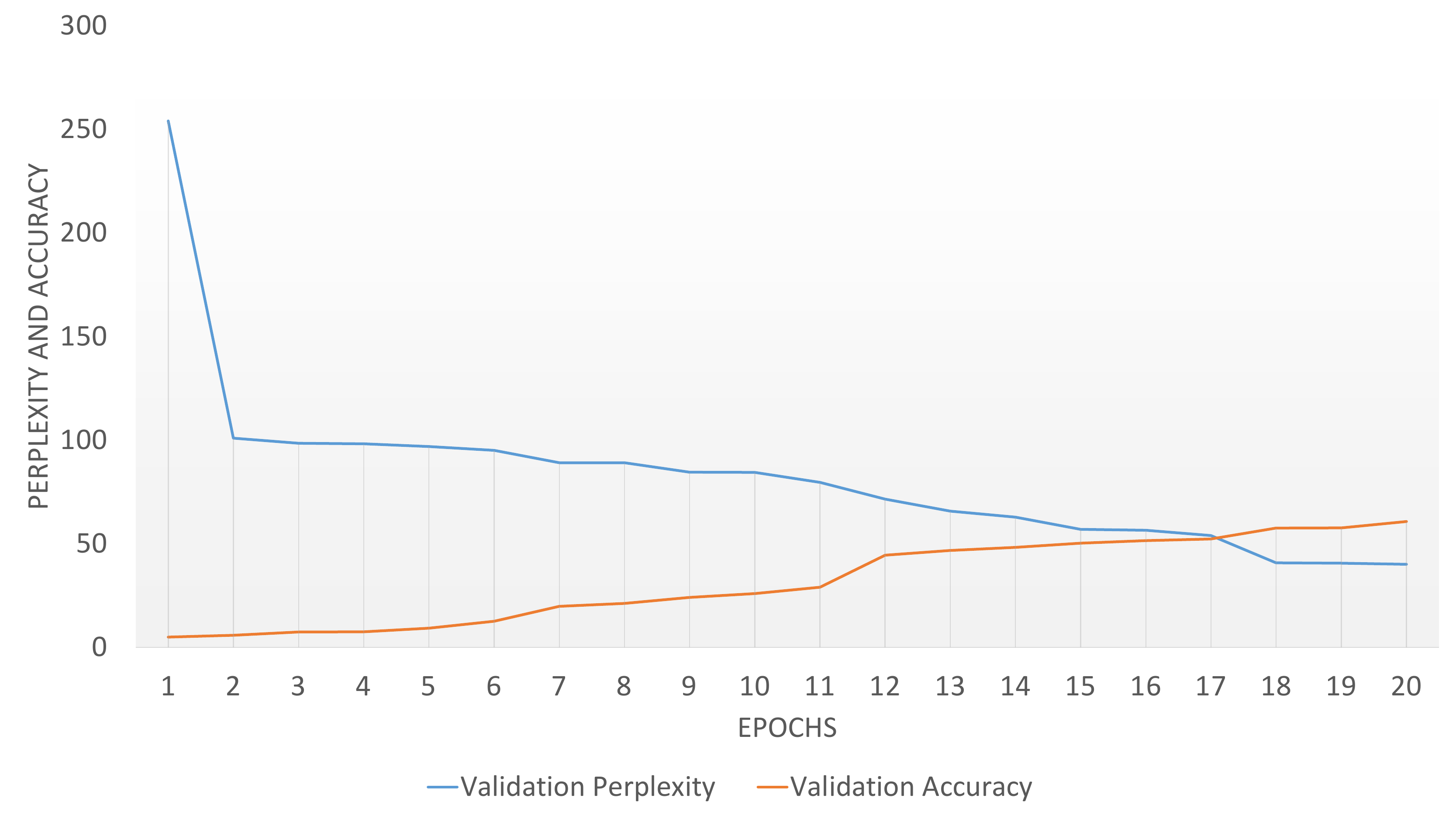}
         \caption{Validation Statistics}
         \label{fig:svs}
     \end{subfigure}
        \caption{Model perplexity vs accuracy statistics for sentences}
        \label{fig:sent_train_val}
\end{figure}

The test is performed for both word level and sentence level question formation. For word level testing of the model performance, the most important word from the sentence is selected and fed into the system. For sentence, the total sentence is passed as an input to the system. For the training and validation statistics model accuracy and perplexity is considered. The accuracy is measured as the total number of words correctly predicted and matched exactly with the labelled data. On the other hand, model perplexity is defined as the errors present in the prediction model.The objective is to always optimize the model by minimizing the perplexity while maximizing the accuracy. The word level training statistics is provided in Figure-\ref{fig:word_train_val} and the sentence level training statistics is provided in Figure-\ref{fig:sent_train_val}. The optimal number of epochs is found to be 20 and hence the model is trained and validated with 20 epochs as represented in the figures. The training and validation statistics confirm that with more length of input there is a higher chance of getting close to the actual data. Since there are many factors associated with text data evaluation like semantic relations, grammar, synonyms etc, which are not evaluated by this measure. To further evaluate the model, metrics based evaluation is also performed.

\subsubsection{Metrics based evaluation}
\label{S:4.5.2}

Evaluation of natural language machine learning models is far from just matching the exact similarity of the prediction to the labelled data. It is due to the fact that the natural language can be framed in various ways but still be correct even if it does not match the original labelled data. This makes it difficult to analyse the models using general metrics Accuracy or F1 score. To overcome the issue, some metrics proposed in the literature such as METEOR and BLEU that are proven to be quite good particularly in this type of task. 

The Bilingual Evaluation Understudy (BLEU), is a score for comparing the predicted text data to a list of reference labelled text data \cite{papineni2002bleu}. It is used to evaluate a wide range of text generation models in natural language processing tasks. Despite of the fact that these evaluation models are not perfect but gives a much reliable evaluation when compared to the general metrics. BLEU is computationally inexpensive and is widely adopted. Further, it is language independent, making the model evaluation more easy to evaluate. In this experiment the medical corpus has a higher frequency of bi-gram data and hence considering those in the evaluation is required. So, we used cumulative n-gram BLEU score with n=1(BLEU-1) and n=2 (BLEU-2). METEOR \cite{banerjee2005meteor} is similar to BLEU in many aspects, but in addition, it is designed to consider synonyms and word stems for more realistic evaluation and it highly correlates with human evaluations. The model prediction statistics with the evaluation metrics for word and sentences are provided in the Table-\ref{tab:model_eval}. From the results, it is clearly visible that a model is benefited always with a sentence rather than a single word due to lack of context. But it does not mean that the model fails to generate an accurate question in spite of this. METEOR score for word tells that the sentence formation is good but BLEU scores tells that it does not match with the labelled data properly. As a side note, the results are produced using GloVes with six billion words and 100 dimensions. Increasing this embedding data and dimension will help by reducing unknown tokens generated and hence will provide much better results. All the experiments performed have been modelled using the same data corpus and embedding for better evaluation.

\begin{table}[H]
\centering
\resizebox{8cm}{!}{%
\begin{tabular}{|c|c|c|c|}
\hline
\textbf{Input Sequence}   & \textbf{METEOR} & \textbf{BLEU-1} & \textbf{BLEU-2} \\ \hline
\textit{Word} & 3.65            & 7.61           & 4.87           \\ \hline
\textit{Sentence} & 5.24               & 9.86               & 4.56              \\ \hline
\end{tabular}%
}
\caption{Model performance evaluation with standard metrics for words and sentence}
\label{tab:model_eval}
\end{table}

\subsubsection{Comparison with the Existing Approaches:}
\label{S:4.5.3}

The state of the art existing approaches for question generation considers a Bi-directional LSTM model with general attention applied at the hidden states with a coverage mechanism for removing redundant information (BiLSTM+GA+C). On the other hand, our model utilizes word and sentence level hierarchical attention mechanism (BiLSTM+HA+C). The experiment is performed separately for word and sentences to evaluate the model performance in both word and sentence level. The results shown in Table-\ref{tab:word_eval} represents the results for the word level model comparison. The hierarchical attention mechanism with word level embedding shows a much higher score for all the evaluation metrics. 

\begin{table}[H]
\centering
\resizebox{8cm}{!}{%
\begin{tabular}{|c|c|c|c|}
\hline
\textbf{Algorithm}   & \textbf{METEOR} & \textbf{BLEU-1} & \textbf{BLEU-2} \\ \hline
\textit{BiLSTM+GA+C} & 2.58            & 5.91           & 2.13           \\ \hline
\textit{BiLSTM+HA+C} & 3.65               & 7.61        & 4.87              \\ \hline
\end{tabular}%
}
\caption{Model comparison with standard metrics for word}
\label{tab:word_eval}
\end{table}

The results shown in Table-\ref{tab:sent_eval} represents the results for the sentence level model comparison. In this experiment, the results are equivalent and does not have much of a difference. Although, the model is similar for the sentence level question generation, the little difference in the scores are mainly due to the fact that many questions can be formed from a single input and hence the the sequences may vary. Overall our model performs much better when considering both the word level and sentence level evaluation.

\begin{table}[H]
\centering
\resizebox{8cm}{!}{%
\begin{tabular}{|c|c|c|c|}
\hline
\textbf{Algorithm}   & \textbf{METEOR} & \textbf{BLEU-1} & \textbf{BLEU-2} \\ \hline
\textit{BiLSTM+GA+C} & 5.26            & 9.67           & 4.51           \\ \hline
\textit{BiLSTM+HA+C} & 5.24            & 9.86           & 4.56             \\ \hline
\end{tabular}%
}
\caption{Model comparison with standard metrics for sentences}
\label{tab:sent_eval}
\end{table}

\subsubsection{Human Evaluations}
\label{S:4.5.4}

Question generation is the process of generation of question either from a single word or a sentence. The testing is based on the labelled data that is sufficient to compare various models in the machine learning field. But, in order to identify the true capability of the question generation model and to provide accurate, error free and context related questions, human evaluations are also considered. We selected 15 volunteers in our lab with a good English background for the offline evaluation of the model and 100 volunteers for an online evaluation making a total of 115 volunteers. For human evaluation of the model performance, three categories are selected with scores between range 0 and 1, as described below:
\begin{enumerate}
    \item \textbf{\textit{Question Selection}} - This parameter is scored according to the systems ability to pick correct questions based on the initial input query provided to the system. It is in fact the relevancy of the question according to the input context.
    \item \textbf{\textit{Question Formation}} - This is the score that defines the quality of the question generation portion of the system. The quality parameters considered are sentence formation and grammatical correctness.
    \item \textbf{\textit{Responsiveness}} - Responsiveness is the systems ability to interact with the user. It is defined as the time taken by the system to respond to the user once an input is provided. This does not include the overall duration of the session, rather it is defined for each interaction made by the system. 
\end{enumerate}
\vspace{0.5cm}
\par The results of human evaluation are provided in the Table-\ref{tab:human_eval}. The human evaluation shows that the system was able to generate semantically correct questions both for word and sentence level inputs with appropriate context. Further, the responsiveness of the system is also acceptable. 

\begin{table}[H]
\scriptsize{
 \begin{tabular}{ |p{2.5cm}||p{2.5cm}|p{2.5cm}|p{2.5cm}|}
  \hline
  \multicolumn{4}{|c|}{Human Evaluation} \\
  \hline
  Parameters & Word Rating & Sentence Rating & Overall Rating\\
  \hline
  Question Selection    &0.5687 &0.6470 &0.7457\\
  Question Formation    &0.3765 &0.7352 &0.7857\\
  Responsiveness        &0.9665 &0.7789 &0.8595\\
  \hline
 \end{tabular}
 \caption{Human evaluation for question generation model}
\label{tab:human_eval}
}
\end{table}

\section{Conclusion and future scope}
\label{S:5}

This paper presents a novel approach of information retrieval using interaction based mechanism and provides enough information regarding the need of similar systems. A totally new framework is built for the identification of questionable entity and with the available evaluation methods, it was identified preforming as intended. Further, addition of word level attention mechanism in the question generation phase has also proven to be effective in improving the overall performance of the model. Comparison with existing techniques for the question generation phase also proves the performance benefit of the proposed model. Further, human evaluation is considered as a qualifier evaluation. Although, the proposed model is able to meet the objectives, there are still possible scopes of improvement such as using a manually trained embedding rather than a pre-trained model.

\bibliography{reference}

\end{document}